\documentstyle[12pt]{article}

\newlength{\dinwidth}
\newlength{\dinmargin}
\def\as{\alpha_{\mbox{\tiny S}}}
\def\alb{\bar\as}
\def\mR{\mu_{\mbox{\tiny R}}}
\def\res{\,{\mbox{\scriptsize jet}}}
\def\om{\omega}

\def\beq{\begin{equation}}
\def\beeq{\begin{eqnarray}}
\def\eeq{\end{equation}}
\def\eeeq{\end{eqnarray}}
\begin{document}
\renewcommand{\thefootnote}{\fnsymbol{footnote}}
\begin{flushright}
Cavendish--HEP--98/15\\
hep-ph/9810286\\
October 1998 \\
\end{flushright}
\begin{center}
\vspace*{2cm}

{\Large \bf\boldmath Jet Rates in Deep Inelastic Scattering at Small $x$
\footnote{Work supported in part by the UK Particle Physics and
Astronomy Research Council and by the EU Fourth Framework Programme
`Training and Mobility of Researchers', Network `Quantum Chromodynamics
and the Deep Structure of Elementary Particles',
contract FMRX-CT98-0194 (DG 12 - MIHT).}}

\vspace*{1cm}
B.R.\ Webber

\vspace*{0.5cm}
Cavendish Laboratory\\
Madingley Road\\
Cambridge CB3 0HE, UK
\end{center}
\vspace*{5cm}
\begin{abstract}
The recent results of Forshaw and Sabio Vera on small-$x$ jet rates to
order $\as^3$ are extended to all orders, for any number of jets.
A simple generating function is obtained. 
\end{abstract}
\newpage

\section{Introduction}
The summation of logarithms of $1/x$ in deep inelastic structure functions
at small values of Bjorken $x$ leads to the Balitskii-Fadin-Kuraev-Lipatov
(BFKL) equation \cite{BFKL,FR}, which in the leading approximation sums
terms of order $[\as\ln(1/x)]^n$. Recently the next-to-leading terms
have also been computed \cite{NLOA,NLOB}.

The usual derivation of the BFKL equation assumes the dominance of
multi-Regge kinematics (i.e.\ strong ordering in the Sudakov variables).
This is valid for the calculation of the totally inclusive structure
functions, in which one sums over all hadronic final states. When
studying the structure of the final states, however, one should take
account of QCD coherence, which effectively imposes an angular ordering
constraint on the emission of soft gluons \cite{C,DKMT,ESW}. The
resulting `CCFM' formulation \cite{C} reduces to the BFKL equation for
the inclusive structure functions, but leads in general to different
exclusive multi-gluon distributions from those expected from
multi-Regge kinematics \cite{M}.

In a recent paper, Forshaw and Sabio Vera \cite{FSV} showed that, in
the leading $\log x$ (LL$x$) approximation, to third order in $\as$,
the rates for emission of fixed numbers of `resolved' final-state gluons,
together with any number of unresolvable ones, are the same in the
multi-Regge (BFKL) and coherent (CCFM) approaches.  Here `resolved'
means having a transverse momentum larger than some fixed value $\mR$,
and the LL$x$ approximation means keeping only terms
that have two large logarithms for each power of $\as$, at least one
of which is $\ln(1/x)$. In this approximation, each resolved gluon
can be equated to a single jet, since to resolve it into more than
one jet would cost extra powers of $\as$ with no corresponding powers
of $\ln(1/x)$.

The present paper extends the work of Forshaw and Sabio Vera to all
orders, for any number of resolved gluons. The BFKL and CCFM formulations
are shown to give the same jet rates in LL$x$ approximation to all
orders. The factorization of collinear singularities is demonstrated,
and a simple generating function for the jet multiplicity distribution
is obtained.

\section{Multi-Regge (BFKL) analysis}
As in Ref.~\cite{FSV}, we start from the unintegrated structure function
of a single gluon, $f(x,k)$, which in the exclusive form of the BFKL
approach satisfies the equation
\beq
f(x,k) = \delta(1-x)\,\delta^2(k) + \alb\int\frac{d^2q}{\pi q^2}
\frac{dz}{z^2}\Delta_R(z,k)\theta(q-\mu)\,f(x/z,q+k)\;.
\eeq
Here $\alb=3\as/\pi$, $k$ is the transverse momentum of the
gluon probed in the deep inelastic scattering,
$q$ is that of an emitted gluon, $\mu$ is a collinear
cutoff (which cancels in the inclusive treatment) and $\Delta_R$ is the
Regge form factor
\beq
\Delta_R(z,k) = \exp\left(-\alb\ln\frac{1}{z}\ln\frac{k^2}{\mu^2}\right)\;.
\eeq
The contribution from emission of $n$ gluons is obtained by iteration,
\beq
f^{(n)}(x,k) = \prod_{i=1}^n\int\frac{d^2q_i}{\pi q_i^2}
\frac{dz_i}{z_i}\alb\Delta_R(z_i,k_i)\theta(q_i-\mu)
\delta(x-x_n)\delta^2(k-k_n)\;,
\eeq
where
\beq
x_i=\prod_{j=1}^i z_j\;,\>\>\>\>k_i=-\sum_{j=1}^i q_j\;.
\eeq
The contribution to the structure function at scale $Q$ is then
obtained by integrating over all $\mu<q_i<Q$:
\beq\label{Fn}
F^{(n)}(x,Q,\mu) = \prod_{i=1}^n\int_{\mu^2}^{Q^2}\frac{dq_i^2}{q_i^2}
\frac{d\phi_i}{2\pi}\frac{dz_i}{z_i}\alb\Delta_R(z_i,k_i)\delta(x-x_n)\;.
\eeq

\subsection{One-jet rate}
Now consider the effect of requiring one emitted gluon, say the $j$th,
to have $q_j>\mR$ while all the others have $q_i<\mR$.  This defines
the contribution of one resolved gluon (plus $n-1$ unresolved),
$F^{(n,1\res)}$:
\beeq
F^{(n,1\res)}(x,Q,\mR,\mu) &=& \sum_{j=1}^n
\int_{\mR^2}^{Q^2}\frac{dq_j^2}{q_j^2}
\frac{d\phi_j}{2\pi}\frac{dz_j}{z_j}\alb\Delta_R(z_j,k_j)\nonumber\\
&\cdot&\prod_{i\neq j}^n\int_{\mu^2}^{\mR^2}\frac{dq_i^2}{q_i^2}
\frac{d\phi_i}{2\pi}\frac{dz_i}{z_i}\alb\Delta_R(z_i,k_i)\delta(x-x_n)\;.
\eeeq
Notice that for $i<j$ the contribution is identical to the $(j-1)$-gluon
contribution to the structure function evaluated at $x=x_{j-1}$ and $Q=\mR$.
On the other hand for $i\geq j$ we have $k_i=q_j$ in leading logarithmic
approximation, and so the $q_i$ integrations become trivial:
\beeq\label{Fn1res}
F^{(n,1\res)}(x,Q,\mR,\mu) &=&
\sum_{j=1}^n \int_0^1 dx_{j-1} F^{(j-1)}(x_{j-1},\mR,\mu)
\int_{\mR^2}^{Q^2}\frac{dq_j^2}{q_j^2}\frac{dz_j}{z_j}
\alb\Delta_R(z_j,q_j)\nonumber\\
&\cdot&\prod_{i=j+1}^n\int\frac{dz_i}{z_i} 2\alb S\Delta_R(z_i,q_j)\delta(x-x_n)
\eeeq
where $S=\ln(\mR/\mu)$.

To carry out the $z_i$ integrations it is convenient to use a Mellin
representation,
\beq
F_\om = \int_0^1 dx\,x^\om F(x)\;,
\eeq
so that
\beeq\label{Fn1resom}
F^{(n,1\res)}_\om(Q,\mR,\mu) &=& \frac{1}{2S}
\sum_{j=1}^n F^{(j-1)}_\om(\mR,\mu)
\int_{\mR^2}^{Q^2}\frac{dq_j^2}{q_j^2}
\left[2\alb S\int\frac{dz}{z} z^\om\Delta_R(z,q_j)\right]^{n-j+1}
\nonumber\\
 &=& \frac{1}{2S}
\sum_{j=1}^n F^{(j-1)}_\om(\mR,\mu)
\int_{\mR^2}^{Q^2}\frac{dq_j^2}{q_j^2}
\left[\frac{2\alb S}{\om + \alb\ln(q_j^2/\mu^2)}\right]^{n-j+1}\;.
\eeeq
Summing over all $j$ and $n$ gives the total one-jet
contribution,
\beq\label{F1res}
F^{(1\res)}_\om(Q,\mR,\mu) = F_\om(\mR,\mu)\,
\int_{\mR}^Q\frac{dq}{q}H_\om(q,\mR)
\eeq
where
\beq\label{Hom}
H_\om(q,\mR) = \frac{2\alb}{\om + 2\alb\ln(q/\mR)}\;,
\eeq
and hence
\beq
F^{(1\res)}_\om(Q,\mR,\mu) = F_\om(\mR,\mu)\,
\ln\left(1+\frac{2\alb}{\om}T\right)
\eeq
where $T=\ln(Q/\mR)$.
The asymptotic behaviour of the structure function is
\beq\label{FommR}
 F_\om(\mR,\mu)\sim\left(\frac{\mR^2}{\mu^2}\right)^{\gamma(\as,\omega)}
\>=\>\exp[2S\gamma(\as,\omega)]
\eeq
where $\gamma(\as,\omega)$ is the Lipatov anomalous dimension:
\beq
\gamma(\as,\omega)= \frac{\alb}{\om}+
2\zeta(3)\left(\frac{\alb}{\om}\right)^4+\ldots
\eeq
Taking the leading term in $\gamma$ gives the result of Ref.~\cite{FSV},
extended to all orders:
\beeq
F^{(1\res)}_\om(Q,\mR,\mu) &=&\exp\left(\frac{2\alb}{\om}S\right)\,
\ln\left(1+\frac{2\alb}{\om}T\right)\nonumber\\
&=&\frac{2\alb}{\om}T +
\left(\frac{2\alb}{\om}\right)^2\left[TS-\frac 1 2 T^2\right]
+\left(\frac{2\alb}{\om}\right)^3\left[\frac 1 3 T^3-\frac 1 2 T^2S
+\frac 1 2 TS^2\right]\nonumber\\
&&+\left(\frac{2\alb}{\om}\right)^4
\left[\frac 1 6 T S^3-\frac 1 4 T^2S^2+\frac 1 3 T^3S
-\frac 1 4 T^4\right]+\cdots\;.
\eeeq
Notice that the collinear-divergent part (the $S$-dependence)
factorizes out, and the fraction of events with one jet is given
by the cutoff-independent function
\beq\label{R1res}
R^{(1\res)}_\om(Q,\mR) = \frac{F^{(1\res)}_\om(Q,\mR,\mu)}{F_\om(Q,\mu)}
\>=\>\ln\left(1+\frac{2\alb}{\om}T\right)\,\exp[-2T\gamma(\as,\omega)]\;.
\eeq

\subsection{Multi-jet rates}
Now suppose we resolve $r$ gluons with transverse momenta $q_j>\mR$.
To leading logarithmic accuracy, the Regge form factors beyond the
first of these have $k_i$ fixed at the largest of the $q_j$'s
resolved so far, and therefore Eq.~(\ref{F1res}) becomes
\beq\label{Frres}
F^{(r\res)}_\om(Q,\mR,\mu) = F_\om(\mR,\mu)\,
\prod_{j=1}^r \int_{\mR}^Q\frac{dq_j}{q_j}H_\om(k_j,\mR)
\eeq
where $k_j = \max_{i\leq j}\{q_i\}$. Introducing $t=\ln(q/\mR)$,
we have the general problem of evaluating
\beq
G^{(r)}_\om(T)\equiv \prod_{j=1}^r \int_0^T
H_\om\left(\max_{i\leq j}\{t_i\}\right)dt_j\;.
\eeq
Defining $G^{(0)}_\om(T)=1$ and introducing the generating function
\beq\label{GuT}
G_\om(u,T)=\sum_{r=0}^\infty u^r G^{(r)}_\om(T)\;,
\eeq
we have
\beeq
G_\om(u,T)&=&\exp\left[\int_0^T dt\left(uH_\om(t)
+u^2tH_\om^2(t)+u^3t^2H_\om^3(t)
+\cdots\right)\right]
\nonumber\\
&=&\exp\left[\int_0^T dt\frac{uH_\om(t)}{1-utH_\om(t)}\right]\;.
\eeeq
From Eq.~(\ref{Hom}) we find in this case
\beq\label{Ht}
H_\om(t) = \frac{2\alb}{\om +2\alb t}\;,\qquad
G_\om(u,T) = \left[1+(1-u)\frac{2\alb}{\om}T\right]^{\frac{u}{1-u}}\;.
\eeq
Thus the $r$-jet rate is given by 
\beq\label{Rrres}
R^{(r\res)}_\om(Q,\mR) = \frac{F^{(r\res)}_\om(Q,\mR,\mu)}{F_\om(Q,\mu)}
\>=\>\frac{1}{r!}
\left.\frac{\partial^r}{\partial u^r}R_\om(u,T)\right|_{u=0}\;,
\eeq
where the jet-rate generating function $R_\om$ is given by
\beq\label{RuT}
R_\om(u,T) = \exp\left(-\frac{2\alb}{\om}T\right)
\left[1+(1-u)\frac{2\alb}{\om}T\right]^{\frac{u}{1-u}}\;.
\eeq

The remarkably simple expression (\ref{RuT}) is the main result
of the present paper. We shall see that the same result is
obtained from the CCFM formulation of small-$x$ dynamics.
After convolution with the measured gluon structure function,
it gives the predicted jet rates in the LL$x$ region
$\ln(1/x)\gg T=\ln(Q/\mR)\gg 1$, to all orders in $\as$.

Expanding to fourth order, we have explicitly
\beeq\label{BFKLRs}
R^{(0\res)}_\om&\simeq&1-\frac{2\alb}{\om}T
+\frac{1}{2} \left(\frac{2\alb}{\om}T\right)^2
-\frac{1}{6}\left(\frac{2\alb}{\om}T\right)^3
+\frac{1}{24}\left(\frac{2\alb}{\om}T\right)^4\nonumber\\
R^{(1\res)}_\om&\simeq&\frac{2\alb}{\om}T
-\frac{3}{2} \left(\frac{2\alb}{\om}T\right)^2
+\frac{4}{3}\left(\frac{2\alb}{\om}T\right)^3
-\left(\frac{2\alb}{\om}T\right)^4\nonumber\\
R^{(2\res)}_\om&\simeq&\left(\frac{2\alb}{\om}T\right)^2
-\frac{13}{6}\left(\frac{2\alb}{\om}T\right)^3
+\frac{23}{8}\left(\frac{2\alb}{\om}T\right)^4\nonumber\\
R^{(3\res)}_\om&\simeq&\left(\frac{2\alb}{\om}T\right)^3
-\frac{35}{12}\left(\frac{2\alb}{\om}T\right)^4\nonumber\\
R^{(4\res)}_\om&\simeq&\left(\frac{2\alb}{\om}T\right)^4\;.
\eeeq
Note that these sum to unity as expected.

From the generating function (\ref{RuT}) we can deduce other
interesting quantities to all orders, for example the mean number
of jets,
\beq
\langle r\rangle =
\left.\frac{\partial}{\partial u}R_\om(u,T)\right|_{u=1}\>=\>
\frac{2\alb}{\om}T +\frac{1}{2} \left(\frac{2\alb}{\om}T\right)^2\;,
\eeq
and the mean square fluctuation in this number,
\beq
\langle r^2\rangle -\langle r\rangle^2 = \frac{2\alb}{\om}T
+\frac{3}{2} \left(\frac{2\alb}{\om}T\right)^2
+\frac{2}{3}\left(\frac{2\alb}{\om}T\right)^3\;.
\eeq
In general, the $p$th central moment of the jet multiplicity
distribution is a polynomial in $\alb T/\om$ of degree $2p-1$,
indicating that the distribution becomes relatively narrow in
the limit of very small $x$ and large $Q/\mR$.  

\section{Analysis including coherence (CCFM)}
Taking account of QCD coherence gives the angular ordering constraint
$q_i>z_{i-1}q_{i-1}$ and in place of Eq.~(\ref{Fn}) one obtains \cite{C}
\beq\label{Fnc}
F^{(n)}(x,Q,\mu) = \prod_{i=1}^n\int_0^{Q^2}\frac{dq_i^2}{q_i^2}
\frac{d\phi_i}{2\pi}\frac{dz_i}{z_i}
\alb\Delta(z_i,k_i,q_i)\theta(q_i-z_{i-1}q_{i-1})\delta(x-x_n)\;,
\eeq
where we set $z_0q_0=\mu$ and $\Delta$ is the `non-Sudakov' form factor
\beq
\Delta(z,k,q) = \exp\left(-\alb\ln\frac{1}{z}\ln\frac{k^2}{z q^2}\right)\;.
\eeq

Corresponding to Eq.~(\ref{Fn1res}) we now have for one resolved emission
\beeq\label{Fn1resc}
F^{(n,1\res)}(x,Q,\mR,\mu) &=& F^{(j-1)}(x_{j-1},\mR,\mu)\,\sum_{j=1}^n 
\int_{\mR^2}^{Q^2}\frac{dq_j^2}{q_j^2}\int_0^{\mR/q_j}\frac{dz_j}{z_j}
\alb\Delta(z_j,q_j,q_j)\nonumber\\
&\cdot&\prod_{i=j+1}^n\int_{z_{i-1}^2q_{i-1}^2}^{\mR^2}\frac{dq_i^2}{q_i^2}
\frac{dz_i}{z_i}\alb\Delta(z_i,q_j,q_i)\delta(x-x_n)\;.
\eeeq
Performing the Mellin transformation and summing over $j$ and $n$,
we obtain in place of Eq.~(\ref{F1res})
\beq\label{F1resc}
F^{(1\res)}_\om(Q,\mR,\mu) = F_\om(\mR,\mu)\,
\int_{\mR}^Q\frac{dq}{q}H_\om(q,q,\mR)
\eeq
where
\beeq\label{Homc}
H_\om(k,q,\mR) &=&2\alb\int_{0}^{\mR/q}\frac{dz}{z} z^\om
\Delta(z,k,q)\,K_\om(z,k,q,\mR)\;,\\
\label{Komc}
K_\om(z,k,q,\mR) &=& 1+2\alb
\int_{zq}^{\mR}\frac{dq'}{q'}\frac{dz'}{z'} {z'}^\om
\Delta(z',k,q')\,K_\om(z',k,q',\mR)\;.
\eeeq

The CCFM treatment predicts the same behaviour as BFKL for the structure
function, and so the factor $F_\om(\mR,\mu)$ in Eq.~(\ref{F1resc})
still has the form (\ref{FommR}).  The explicit solution of
Eq.~(\ref{Komc}) is
\beq\label{Komsol}
K_\om(z,k,q,\mR) = \frac{\om+2\alb\ln(k/zq)}{\om+2\alb\ln(k/\mR)}\;.
\eeq
Introducing $s=\ln(k/\mR)$, $t=\ln(q/\mR)$, we can write the
expression in Eq.~(\ref{Homc}) as
\beq\label{Hst}
H_\om(s,t)=\frac{2\alb}{\om+2\alb s}\exp\left(\alb(t^2-2st)-\om t\right)\;.
\eeq
The exponential factor contributes only sub-leading corrections,
i.e.\ terms with fewer factors of $\ln(1/x)$ than of $\as$,
which we are neglecting. Substituting in Eq.~(\ref{F1resc}), we therefore
find that to the same precision the CCFM result for the one-jet
contribution is equal to the BFKL prediction (\ref{F1res}).

To find the multi-jet fractions we use Eq.~(\ref{GuT}) with
\beq
G^{(r)}_\om(T) = \prod_{j=1}^r \int_0^T
H_\om\left(\max_{i\leq j}\{t_i\},t_j\right)dt_j\;,
\eeq
$H_\om$ now being a function of two variables, given by Eq.~(\ref{Hst}).
However, $H_\om$ depends only on its first argument, apart from the
negligible exponential factor.  Therefore the CCFM rates for
more than one jet also have the BFKL values, given by Eq.~(\ref{RuT}).
This completes the all-orders extension of the results of Ref.~\cite{FSV}.

\section{Conclusions}
The above results show that the multi-jet rates in deep inelastic
scattering at small $x$, as defined here, are sufficiently inclusive
quantities to be insensitive to the differences between the BFKL
and CCFM formulations of small-$x$ dynamics at the leading-logarithmic
level. In both cases they are given by the simple generating function
(\ref{RuT}). Differences would be expected at the sub-leading level and
in more differential quantities such as multi-jet rapidity
correlations \cite{CMW}.

Bearing in mind the importance of sub-leading corrections to
structure functions at small $x$ \cite{NLOA,NLOB}, one would not
expect these result to be of direct phenomenological relevance,
although they may provide a useful cross-check on the results of
numerical simulations of small-$x$ final states \cite{K,SMC,SO}.

\section*{Acknowledgements}
I am grateful to the CERN Theory Division for hospitality while this
work was performed, and to G.\ Marchesini and S.\ Catani for many helpful
discussions.

\end{document}